\documentclass[journal, ]{IEEEtran}
\title{Artificial-Noise Alignment for Secure Multicast using Multiple Antennas}
\usepackage{amssymb}
\usepackage{amsmath}
\usepackage{verbatim}
\usepackage{anysize}
\usepackage{verbatim} 
\usepackage{cite}
\usepackage{graphicx}
\usepackage{epsfig}
\usepackage{psfrag}

\DeclareMathAlphabet{\mathbsf}{OT1}{cmss}{bx}{n}
\DeclareMathAlphabet{\mathssf}{OT1}{cmss}{m}{sl}

\newcommand{\bh}{{\mathbf{h}}}
\newcommand{\bg}{{\mathbf{g}}}

\newcommand{\bx}{{\mathbf{x}}}

\newcommand{\rvy}{{\mathssf{y}}}
\newcommand{\rvz}{{\mathssf{z}}}

\newcommand{\rvv}{{\mathssf{v}}}
\newcommand{\rvw}{{\mathssf{w}}}

\newcommand{\cN}{{\mathcal{N}}}
\newcommand{\cT}{{\mathcal{T}}}
\newcommand{\cA}{{\mathcal{A}}}

\newcommand{\bv}{{\mathbf{v}}}

\newcommand{\bb}{{\mathbf{b}}}
\newcommand{\bht}{{\tilde{\bh}}}
\newcommand{\bgt}{{\tilde{\bg}}}
\newcommand{\bgh}{{\hat{\bg}}}
\newcommand{\bhh}{{\hat{\bh}}}

\newcommand{\bu}{{\mathbf{u}}}
\newcommand{\cC}{{\mathcal{C}}}

\newcommand{\eps}{\varepsilon}
\newcommand{\al}{\alpha}
\newcommand{\bal}{{\boldsymbol{\al}}}
\newcommand{\g}{\gamma}

\newtheorem{prop}{Proposition}

\author{Ashish Khist
        and Dongye Zhang
\thanks{This work was supported by an NSERC Discovery Grant }%
\thanks{A. Khisti and D.~Zhang are with the Dept.\ Electrical and Computer
  Engineering, University of Toronto, ON, Canada M5S 3G4.  (Email:
  akhisti@comm.utoronto.ca). }}

\begin{document}

\maketitle

\begin{abstract}

We propose an artificial-noise alignment scheme for multicasting a common-confidential message to a group of receivers. Our scheme transmits a superposition of information and noise symbols. The noise symbols are aligned at each legitimate receiver and hence the information symbols can be decoded. In contrast, the noise symbols completely mask the information symbols at the eavesdroppers. Our proposed scheme does not require the knowledge of the eavesdropper's channel gains at the transmitter for alignment, yet it achieves the best-known lower bound on the secure degrees of freedom. Our scheme is also a natural generalization of the approach of transmitting artificial noise in the null-space of the legitimate receiver's channel, previously proposed in the literature.
\end{abstract}

\section{Introduction}
Multiple antennas provide a promising approach for enhancing confidentiality of  messages at the physical layer.  A natural technique using multiple antennas is artificial-noise transmission~\cite{negi}. We transmit an information
message by beamforming in the direction of the legitimate receiver and superimpose a noise signal in a direction orthogonal to the legitimate receiver.
This way any eavesdropper, whose channel vector has a component along the noise vector gets jammed by the noise signal.
Unfortunately, such an approach does not scale when we need to multicast a common message to a large number of legitimate
receivers. If the number of receivers is larger than the number of transmit antennas, we cannot find a vector that is simultaneously in the null space
of all receivers.  The main purpose of this note is to show that in such situations, one can transmit a superposition of signal and noise.
If the transmitter has $M$ antennas,  one should align the noise symbols at each legitimate receiver so that together they occupy only $\frac{1}{M}$ degrees of freedom, and the information
symbols can occupy the remaining $1-\frac{1}{M}$ degrees of freedom.

In related works, the multi-antenna compound wiretap channel was introduced in~\cite{liangKramer:08} where a common message needs to be
transmitted to a group of legitimate receivers and needs to be kept confidential from a group of eavesdroppers.
An interference alignment scheme for this setup was proposed in~\cite{Khisti11}. The interference alignment scheme in~\cite{Khisti11}
however requires  knowledge of the eavesdropper channel gains.
Using the real-interference alignment approach in~\cite{realInterf1, realInterf2, compoundMIMO1}, it aligns the information
symbols at each eavesdropping receiver so that they only occupy $\frac{1}{M}$ degrees of freedom at each eavesdropper.  In contrast, our proposed scheme only requires knowledge of the channel gains of the legitimate receivers for alignment, yet achieves the same degrees of
freedom as in~\cite{Khisti11}.

\section{Channel Model}
We consider a compound multi-antenna wiretap channel that consists of one transmitter with $M$ antennas,
and a group of $J_1$ legitimate receivers, each with one antenna, and a group of $J_2$ eavesdroppers each with one antenna.
The resulting channel model can be expressed as
\begin{equation}
\begin{aligned}
\rvy_j &= \bh_j^T \bx + \rvv_j, \qquad j=1,\ldots, J_1 \\
\rvz_k &= \bg_k^T  \bx + \rvw_k,\qquad k = 1,\ldots, J_2,
\end{aligned}\label{eq:model}
\end{equation}where the transmitted signal vector $\bx\in \mathbb{R}^M$ is required to satisfy the average power constraint $E[||\bx||^2]\le P$,
and the additive noise variables $\rvv_j$ and $\rvw_k$ are independent identically distributed (i.i.d) AWGN noise variables,  distributed $\cN(0,1)$.
In our model we assume that all the elements of the vectors $\bh_j, \bg_k \in \mathbb{R}^M$ are rationally independent; such a condition is satisfied with probability $1$
if the channel gains are sampled independently from any continuous valued distribution. We will assume that the channel gains $\bh_j$ of the legitimate
receivers are perfectly known to the transmitter. The channel gains $\bg_k$ are not known to the transmitter. However when selecting the rate of the code, we will
assume that an upper bound  \begin{equation}\max_k ||\bg_k||^2 \le c\label{eq:c}\end{equation} is known to the transmitter.
Furthermore, we only consider the case that the channel coefficients remain fixed for the entire duration of communication.

We transmit a single common message $\rvw$ to all the $J_1$ legitimate receivers. A rate $R$ is achievable if there exists a sequence of length $n$
wiretap codes such that the error probability at each legitimate receiver goes to zero as $n\rightarrow\infty$ and the leakage rate $\frac{1}{n} I(\rvw; \rvz_k^n)$ also approaches zero
as $n\rightarrow\infty$ for each $k=1,\ldots, J_2$. Of particular interest is the achievable degrees of freedom i.e., $d = \lim_{P\rightarrow\infty}\frac{R}{\frac{1}{2}\log_2 P}$.
As remarked earlier, the result in~\cite{Khisti11} assumes a complete knowledge of channel gains of the eavesdropper and proposes a signal alignment scheme that achieves $d = 1-\frac{1}{M}$.
Our main result is that the same degrees of freedom can be achieved using a noise alignment scheme that aligns artificial-noise in the direction of legitimate receivers. Such a scheme
has the  advantage that it does not need any information about the eavesdropper's channel in the alignment process.

\begin{prop}
An achievable degree-of-freedom for the compound multi-antenna wiretap channel with $M$ transmit antennas, where the channel gains of the legitimate receivers are perfectly known,
whereas the channel gains of the eavesdropper are not known, except for $c$ (c.f.~\eqref{eq:c}), is given by $d = \left(1-\frac{1}{M}\right)$.
\end{prop}

\section{Artificial-Noise Alignment}

Our transmission scheme consists of sending fictitious (noise) messages in addition to the information message $\rvw$.
Through an appropriate choice of a precoder, we align the noise symbols at each legitimate receiver while the noise symbols are not aligned at any eavesdropper.  This enables the legitimate receiver
to decode the information message, whereas it is completely masked by the fictitious messages at the eavesdropper.

We begin by defining the precoding sets as follows. Let $N$ be a sufficiently large integer and let
\begin{align}\label{eq:RI-Ct}
\cT &= \left\{\prod_{j=1}^{J_1}\prod_{i=1}^M h_{ji}^{\alpha_{ji}} \big|~  \alpha_{ji} \in \{0,\ldots, N-1\}\right\},\\
\cA &= \left\{\prod_{j=1}^{J_1}\prod_{i=1}^M h_{ji}^{\alpha_{ji}} \big|~  \alpha_{ji} \in \{0,\ldots, N\}\right\},\label{eq:RI-A}
\end{align}where $h_{ji}$ denotes the channel gain between the $i$-th transmitter antenna and the $j$-th legitimate receiver.
Note that each selection of the tuple $\{\alpha_{ji}\} \in \{0,\ldots, N-1\}^{MJ_1}$ results in a different element of $\cT$. There are a total of $L = N^{MJ_1}$ elements in $\cT$, and $L' = (N+1)^{MJ_1}$ elements in $\cA$. Let $\bv \in \mathbb{R}^{L}$ consists of all elements in the set $\cT$, and let
\begin{equation}
V = \left[\begin{array}{cccc} \bv^T & 0 &\cdots & 0 \\ 0 & \bv^T & \cdots & 0 \\ \vdots &   & \ddots & \vdots \\ 0 & 0 &\cdots & \bv^T \end{array}\right]\in \mathbb{R}^{M \times ML}.
\label{V}\end{equation}
We let our signal constellation be
\begin{equation}
\cC = a\left\{-Q,-Q+1,\ldots, Q-1, Q\right\}\label{eq:dsPAM}
\end{equation} where the value of $a$ and $Q$ will be defined in the sequel.
We let the transmit vector be given as
\begin{equation}
\bx = \bu (\boldsymbol{\alpha}^T\bb_1) + V \bb_2 \label{x}
\end{equation}
where $\bb_1 \in \cC^{KL}$ is the vector of information symbols  and $\bb_2\in \cC^{ML}$ is the vector of fictitious noise symbols. All the symbols are
uniformly distributed over $\cC$. Furthermore $\bu \in \mathbb{R}^{M}$  and $\boldsymbol{\alpha} \in \mathbb{R}^{KL}$ are vectors whose elements are mutually  independent and also independent of all the elements in $h_{ji}$.  The output at each legitimate receiver can be expressed as
\begin{equation}
\rvy_j = \bh_j^T \bu \boldsymbol{\alpha}^T\bb_1 + \bh_j^T V \bb_2  +\rvv_j, \qquad j=1,\ldots, J_1.
\end{equation}
Let $\bhh_j =\left( \bh_j^T \bu \boldsymbol{\alpha}^T \right) ^T,$ {so }  that $\bhh_j \in\mathbb{R}^{KL}$. Notice that the elements of $\bhh_j$ are rationally independent and independent of the elements in $\cA$. We let \begin{equation}\bht_j^T=\bh_j^T V = \left[h_{j1}\bv^T,\ldots,h_{jM}\bv^T\right].\label{eq:bgt}\end{equation}Thus, $\bht_j$ is a length $ML$ vector whose elements belong to the set $\cA$ in~\eqref{eq:RI-A}.  Since all elements of $\bht_j$ belong to $\cA$  we can also express it as~\cite[Lemma 2]{Khisti11}
\begin{equation}
\bht_j^T = \bht^T T_j,
\end{equation}
where $\bht \in \mathbb{R}^{L'}$ is a vector consisting of all the elements in $\cA$ and $T_j \in \mathbb{R}^{L' \times ML}$ is a matrix for which every column has exactly one element that equals $1$, and the remaining elements are zero. Furthermore no more than $M$ entries in each row of $T$ are non-zero.  The output at each legitimate receiver can be simplified as
\begin{equation} \label{eq:yj}
\rvy_j = \bhh_j^T \bb_1 + \bht^T T_j\bb_2  +\rvv_j, \qquad j=1,\ldots, J_1
\end{equation}
Note that the elements of $\bhh_j$ and $\bht$ are rationally independent. The output at each eavesdropper can be expressed as
\begin{equation}\label{eq:zk}
\rvz_k = (\bg_k^T \bu) \boldsymbol{\alpha}^T\bb_1 + \bg_k^T V \bb_2  +\rvw_k, \qquad k=1,\ldots, J_2.
\end{equation}
We let \begin{equation}\bgt_k^T=\bg_k^T V = \left[g_{k1}\bv^T,\ldots,g_{kM}\bv^T\right].\end{equation}
Eq.~\eqref{eq:zk} then reduces to
\begin{equation}
\rvz_k = \bgh_k^T \bb_1 + \bgt_k^T  \bb_2  +\rvw_k, \qquad k=1,\ldots, J_2.
\end{equation}
where $\bgh_k=\left( \bg_k^T \bu \boldsymbol{\alpha}^T \right) ^T \in \mathbb{R}^{KL}$.

Since the elements of $\bu$, $\boldsymbol{\alpha}$, $\bh_j$ and $\bg_k$  are rationally independent, it follows that all the elements of $\bgh_k$ and $\bgt_k$ are rationally independent.

In the next section, we develop an expression for an achievable rate for the proposed coding scheme and propose  the value for $K$ that maximizes the
degrees of freedom.

\section{Secrecy rate}
Since the elements of the vectors $\bhh_j \in \mathbb{R}^{KL}$ and $\bht \in \mathbb{R}^{L'}$ in~\eqref{eq:yj} are rationally independent, an appropriate scaling of constellation parameters exist that achieves the rate close to capacity. In particular we select
\begin{equation}
Q =  P^{\frac{1-\eps}{2(KL+L'+\eps)}},\qquad a =\g\frac{P^{\frac{1}{2}}}{Q}\label{eq:Qa}
\end{equation}
where $\g^2 = \frac{1}{M(||\bhh_j||^2 + ||\bht||^2)}$ is a normalizing constant which only depends on the channel gains and predefined $\bu$ and $\bal$.
An achievable secrecy rate for the compound wiretap channel model is~\cite{liangKramer:08}
\begin{align}
R &= \max_{p_{\text{}\bb_1,\bx}}\left\{\min_j I(\bb_1;\rvy_j) - \max_k I(\bb_1;\rvz_k) \label{eq:Rdiff}\right\}.
\end{align}

To compute~\eqref{eq:Rdiff} we note that
\begin{equation}
I(\bb_1;\rvy_j) - I(\bb_1;\rvz_k) = H(\bb_1) - H(\bb_1|\rvy_j) - I(\bb_1;\rvz_k) \label{eq:RdiffSimp}
\end{equation}
and bound each of the three terms. Since all elements of $\bb_1 \in \mathbb{R}^{KL}$ are uniformly distributed over the constellation $\cC$ it follows that
\begin{equation}
\label{eq:termb1}
H(\bb_1) = KL \log_2(2Q+1)\\
\ge \frac{1}{2}\frac{KL(1-\eps)}{KL+L'+\eps}\log_2 P
\end{equation}

Next since the vectors $\bhh_j \in \mathbb{R}^{KL}$ and $\bht \in \mathbb{R}^{L'}$ in~\eqref{eq:yj} consist of rationally independent channel gains, we have (c.f.~\cite{Khisti11}) that the error probability at legitimate receivers
\begin{equation}
\Pr(e) \le \exp\left(-\eta P^\eps\right) = o_P(1),\label{eq:pe}
\end{equation}
where $\eta>0 $  that depends on channel gains but does not depend on $P$ and $o_P(1) \rightarrow 0$ as $P \rightarrow \infty$. Thus from Fano's inequality we have,\begin{align}
H(\bb_1|\rvy_j) &\le 1 + \Pr(e) H(\bb_1) \notag \\
&= 1 + \Pr(e) KL \log_2(2Q+1)\notag \\
&= 1 + o_P(1)\log_2P.\label{eq:termb1yj}
\end{align}

Note that since the channel gains in $\bgt_k$
are rationally independent, $\bb_2$ can be decoded with high probability from $\bgt_k^T  \bb_2  +\rvw_k$ on the condition that
\begin{equation}
KL + L' \ge ML. \label{eq:condition}
\end{equation}
Hence, we have
\begin{align}
&I(\bb_1;\rvz_k)=I(\bb_1,\bb_2;\rvz_k)-I(\bb_2;\rvz_k|\bb_1) \notag \\
&=I(\bx;\rvz_k)-H(\bb_2)+H(\bb_2|\rvz_k,\bb_1)  \notag \\
&=I(\bx;\rvz_k)-H(\bb_2)+ o_P(1) \label{eq:termb1zkStep1}\\
&\le \frac{1}{2} \log_2 (c P+1) -H(\bb_2)+ o_P(1) \label{eq:cbnd}\\
&\le \frac{1}{2} \log_2 (c P+1) - \frac{1}{2}\frac{ML(1-\eps)}{KL+L'+\eps}\log_2 P +o_P(1) \label{eq:termb1zk}
\end{align}
where~\eqref{eq:termb1zkStep1} follows based on the condition~\eqref{eq:condition} and~\eqref{eq:cbnd}
follows from the fact that $I(\bx;\rvz_k) \le \log\left(1 + ||\bg_k||^2 P\right) \le \log(1 + cP)$ where $c$ is defined
in~\eqref{eq:c}. Substitute~\eqref{eq:termb1},~\eqref{eq:termb1yj} and~\eqref{eq:termb1zk} into~\eqref{eq:RdiffSimp}, we can achieve the lower bond on the secrecy rate
\begin{align}
R &= I(\bb_1;\rvy_j) - I(\bb_1;\rvz_k) \notag \\
&= H(\bb_1) - H(\bb_1|\rvy_j) - I(\bb_1;\rvz_k) \notag \\
&\ge \frac{1}{2}\frac{KL(1-\eps)}{KL+L'+\eps}\log_2 P - \left( 1+o_P(1)\log_2 P\right) \notag- \\ & \left(\frac{1}{2} \log_2 (cP+1) - \frac{1}{2}\frac{ML(1-\eps)}{KL+L'+\eps}\log_2 P +o_P(1)\right) \notag\\
&\ge \frac{1}{2} \left( \frac{(K+M)L(1-\eps)}{KL+L'+\eps}-1-o_P(1)\right)\log_2 P \notag
\end{align}

Finally, we have the secure degree of freedom (d.o.f)
\begin{align}
d=\lim_{P\rightarrow\infty}\frac{R}{\frac{1}{2}\log P} &= \frac{(K+M)L(1-\eps)}{KL+L'+\eps}-1 \label{eq:dof}
\end{align}

We need to select $K$ to maximize $d$ given the constraint in~\eqref{eq:condition}. We select
\begin{align}
K = M - \frac{L'}{L}
= M - \frac{N^{MJ_1}}{(N+1)^{MJ_1}}
\end{align} By selecting $N$ sufficiently large, $K \rightarrow M-1$ and $\eps$ can be selected to be sufficiently close to zero, the secure d.o.f in~\eqref{eq:dof} can be made arbitrarily close to $1-\frac{1}{M}$.

\vspace{-1em}

\section{Conclusion}
We propose the use of artificial-noise alignment  for transmitting a confidential message using a multi-antenna transmitter.
The proposed scheme transmits a superposition of information and noise symbols. It simultaneously aligns the noise symbols
at all intended users so that the message symbols can be decoded by these receivers. In contrast, the message symbols are completely masked by noise
symbols at the eavesdroppers. Our alignment scheme does not require the knowledge of eavesdropper channel gains and the codebook construction
only requires a bound on the SNR of the eavesdroppers.

\end{document}